\documentclass[preprint,amsmath,amssymb,aps,prb]{revtex4-1}
\usepackage{epsfig}
\usepackage{graphicx}
\usepackage{dcolumn}
\usepackage{bm}
\usepackage{indentfirst}
\usepackage{latexsym}
\begin{document}
 
\title{Electron Transport in MoWSeS Monolayers in Presence of an External Electric Field}
\author{Nourdine Zibouche,$^{1,2}$ Pier Philipsen,$^2$ Thomas Heine$^1$ and Agnieszka Kuc$^1$}
\email{a.kuc@jacobs-university.de}
\affiliation{$^1$School of Engineering and Science, Jacobs University Bremen, Campus Ring 1, 28759 Bremen, Germany\\
$^2$Scientific Computing \& Modelling NV, De Boelelaan 1083, 1081 HV Amsterdam, The Netherlands}
 
\date{}
\begin{abstract}
The influence of an external electric field on single-layer transition-metal dichalcogenides TX$_2$ with T = Mo, W and X = S, Se (MoWSeS) have been investigated by means of density-functional theory within two-dimensional periodic boundary conditions under consideration of relativistic effects including the spin-orbit interactions. Our results show that the external field modifies the band structure of the monolayers, in particular the conduction band. This modification has, however, very little influence on the band gap and effective masses of holes and electrons at the $K$ point, and also the spin-orbit splitting of these monolayers is almost unaffected. Our results indicate a remarkable stability of the electronic properties of TX$_2$ monolayers with respect to gate voltages. A reduction of the electronic band gap is observed only starting from field strengths of 2.0 V \AA$^{-1}$ (3.5 V \AA$^{-1}$) for selenides (sulphides), and the transition to a metallic phase would occur at fields of 4.5 \AA$^{-1}$ (6.5 \AA$^{-1}$).
\end{abstract}
 
\maketitle

Monolayered transition-metal dichalcogenides, such as TX$_2$ with T = Mo, W and X = S, Se (MoWSeS), have emerged as novel attractive two dimensional (2D) materials for potential applications in nano- and optoelectronic devices.\cite{Wang2012a} 
TX$_2$ of 2$H$ symmetry are hexagonal systems with one layer of transition-metal atoms sandwiched between two layers of chalcogen atoms.
In the bulk form, the adjacent sheets are held together via weak interlayer forces, thus allowing easy and fast mechanical or chemical exfoliation to the monolayered forms.\cite{Mak2010, Radisavljevic2011, Coleman2011}
Recently, Kis and co-workers have made a substantial breakthrough by using MoS$_2$ monolayers to fabricate field-effect transistors,\cite{Radisavljevic2011} integrated circuits\cite{Radisavljevic2011a}, amplifiers\cite{Radisavljevic2012} and photodetectors.\cite{Lopez-Sanchez2013}

Semiconducting MoWSeS materials undergo the indirect to direct band gap transition when thinned to the monolayer limit.\cite{Splendiani2010, Mak2010, Kuc2011}
At this limit, the inversion symmetry is lost what causes a giant spin-orbit-induced band splitting of 100 meV for MoS$_2$ monolayer from Raman experiments,\cite{Sun2013} 148 meV for MoS$_2$ monolayer up to 456 meV for WTe$_2$ from first principles calculations.\cite{Zhu2011}
The electronic properties of these 2D systems can be further tuned by mechanical distortions, such as tensile strain.
For example, by applying a small mechanical strain of about 1\% to the MoS$_2$ monolayer, the band gap shifts from direct to indirect, and for larger deformations a semiconductor-metal transition occurs.\cite{Scalise2012, Lu2012, Johari2012, Ghorbani2013, Ghorbani2013a}

By means of further theoretical studies it has been reported that applying an external electric field to a rippled MoS$_2$ monolayer\cite{Qi2013} or an armchair MoS$_2$ nanoribbon\cite{QuYue2012} reduces the band gap and causes severe changes in the electronic structure.
Ramasubramaniam and co-workers\cite{Ramasubramaniam2011} have studied the effect of the perpendicular external electric field applied to TX$_2$ bilayers.
Their results, obtained via first principles based plane wave calculations, indicate that the band gap decreases linearly with the external electric field, resulting in a semiconductor-metal transition in the range of relatively small electric field of 200-300 mV \AA$^{-1}$.
On the other hand, Liu et al.\cite{QLiu2012} have reassessed the change of electronic structure of a MoS$_2$ bilayer in the presence of a perpendicular electric field, considering different stacking configurations of molybdenum and sulphur atoms in the 2D layers.
They found that the electric field strength at which the band gap closes is significantly higher, between 1.0 and 1.5 V \AA$^{-1}$.
The strongly underestimated values of Ramasubramaniam et al.\cite{Ramasubramaniam2011} are caused by applying inappropriate constrains to the symmetry of the bilayer structures.
In addition, it has been reported that the band gaps of TX$_2$ monolayers are insensitive to perpendicular external fields in this range of strength.\cite{Ramasubramaniam2011, QLiu2012}

In this study, we have calculated the effect of a perpendicular external electric field on the electronic structure of MoWSeS monolayers from first-principles explicitly considering spin-orbit interactions.
In contrast to previous studies, we applied 2D periodic boundary conditions within the TX$_2$  layers and thus avoid spurious periodicity in the applied electric field and resulting polarization normal to the monolayers.
Our results show that the electronic structure of MoWSeS monolayers show that the electronic structure is not affected by electric fields that are common in electronic devices.
First changes are observed for the conduction band, which changes more strongly than the related electronic band gaps and charge carrier mobilities.
For all materials, we observe a transition from direct to indirect band gap for field strengths of about 2.0 V \AA$^{-1}$.
Electronic band gap and effective masses of electrons and holes stay almost unaffected at the $K$ point for field strengths below 2.0 V \AA$^{-1}$ (3.5 V \AA$^{-1}$) for selenide (sulphide) materials.
The band splitting in the monolayers, a result of spin-orbit coupling, remains unaffected for the whole range of electric fields studied in the present work.
As the field does not affect the electron and hole effective masses, our calculations suggest that the electronic transport properties remain unchanged if the monolayers are subjected to an even excessively high gate voltage.
 
We have studied the changes in the electronic band structures of TX$_2$ monolayers (T = Mo, W; X = S, Se) with respect to the applied perpendicular external electric field.
All calculations were carried out using density functional theory (DFT) at the GGA-PBE\cite{PBE} level as implemented in the ADF-BAND software.\cite{BAND1, BAND}
Mixed numerical and Slater-type orbitals with valence triple-zeta quality and one polarization function (TZP) were adopted for all the atoms, together with a small frozen core.
The structures were fully optimized (atomic positions and lattice vectors) and the maximum gradient threshold was set to $10^{-4}$ Hartree \AA$^{-1}$.
Relativistic effects were taken into account for the optimization procedure by employing the scalar Zero Order Regular Approximation (ZORA).\cite{Pier1997}
Electronic band structure calculations were performed on the optimized structures employing the spin-orbit coupling (SOC) and an external electric field normal to the basal plane of the monolayers.
The $k$-point mesh over the first Brillouin zone was sampled according to the Wiesenekker-Baerends scheme,\cite{Zaharioudakis1991} where the $k$-space integration parameter is set to 5, leading to 15 $k$-points in the irreducible wedge.

\begin{figure}[h!]
\begin{center}
\includegraphics[scale=0.25,clip]{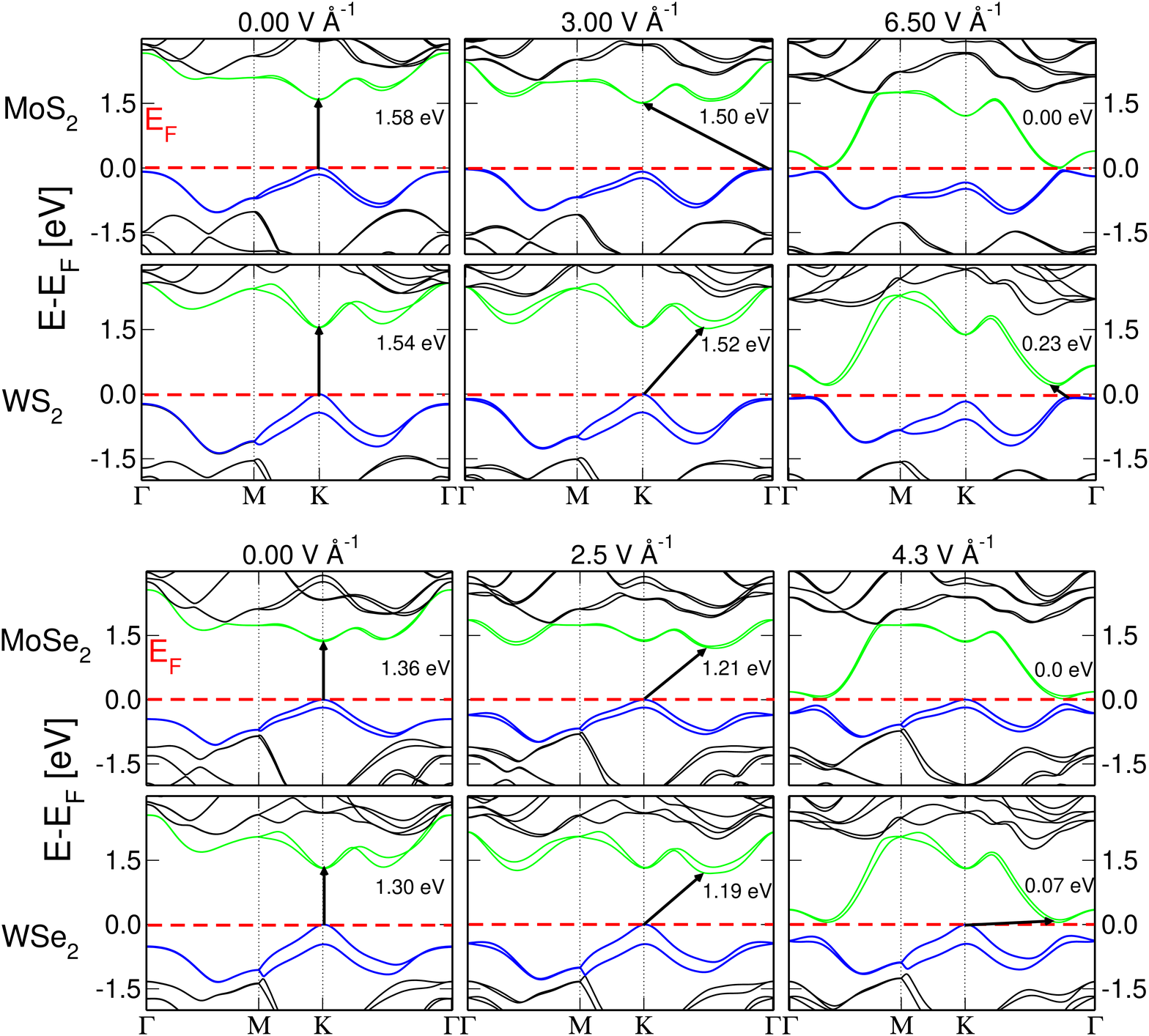}
\caption{\label{fig:BS_S} Electronic band structures of TX$_2$ systems with respect to the external electric field. The top of valence and bottom of conduction band are highlighted with blue and green, respectively. The Fermi level (E$_F$) is shifted to the top of valence band. The values of fundamental band gaps are given.}
\end{center}
\end{figure}

We have applied an external electric field normal to the basal planes of MoWSeS monolayers.
The range of the electric field strengths considered in the present studies is 0.0--7.5 V \AA$^{-1}$.
Fig.~\ref{fig:BS_S} shows the change in the band structures of the 2D systems for selected external field strengths.
In their equilibrium structures, all systems are direct band gap semiconductors at the $K$ point.
Application of the electric field changes the position of the conduction band minimum (CBM) to the 2/3 position between $K$ and $\Gamma$ and the systems become indirect band gap materials.
The only exception is found for the MoS$_2$ monolayer.
Here, the transition between $\Gamma$ and $K$ is very similar in energy.
Moreover, sulphide systems get metallic only at larger field strength compared to the corresponding selenide materials, at about 6.5 versus 4.5 V \AA$^{-1}$.
We observe the so-called Stark effect, resulting in a shift of the bands and in a change of the band structure in presence of an external electric field, especially in the conduction region.
The Stark effect is most pronounced for the WX$_2$ monolayers.

\begin{figure}[h!]
\begin{center}
\includegraphics[scale=0.2,clip]{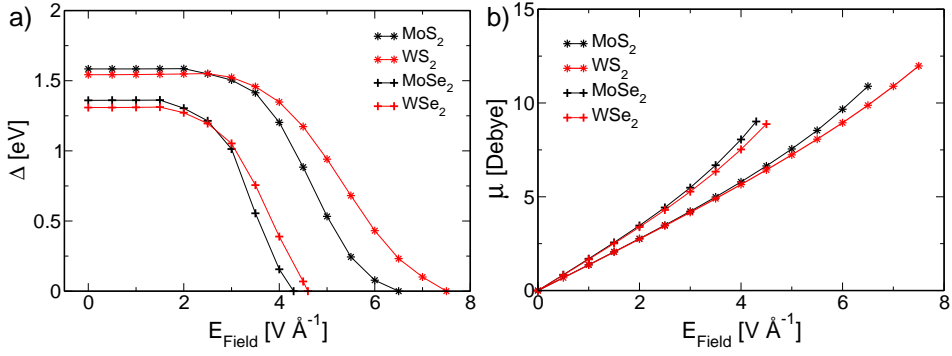}
\caption{\label{fig:gaps} Calculated band gaps (a) and total dipole moments (b) versus external electric field of MoWSeS monolayers.}
\end{center}
\end{figure}

Fig.~\ref{fig:gaps} shows the band gap and the dipole moment evolution of MoWSeS monolayers with respect to the field strength.
The band gaps stay almost constant up to 3.5 and 2.0 V \AA$^{-1}$ for sulphides and selenides, respectively.
At the same time the dipole moments increase linearly with external field strength, even in the regions where the band gaps are unaffected.
Above a critical field strength, the decrease in the band gap is more rapid for the Mo-based systems compared with the W-counterparts.
As the strength of the electric field increases, the polarization also increases and hence the dipole moment changes deviate from linearity, leading to the band gaps closure for critical electric fields.

\begin{figure}[h!]
\begin{center}
\includegraphics[scale=0.28,clip]{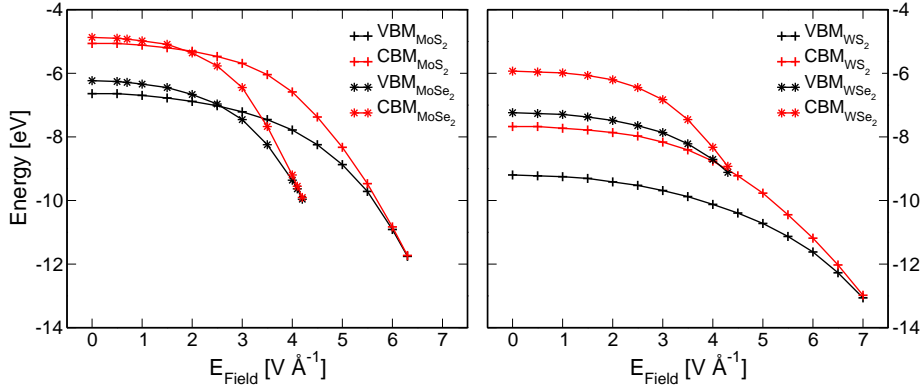}
\caption{\label{fig:bm} Energies of the valence band maximum (VBM) and conduction band minimum (CBM) versus external electric field of MoWSeS monolayers.}
\end{center}
\end{figure}

The changes in the energies of the valence band maximum (VBM) and CBM with the electric field are shown in Fig.~\ref{fig:bm}.
Slow reduction in the energy of both extrema are observed even for weak fields, however, above the critical values of electric field 2 V \AA$^{-1}$ (3.5 V \AA$^{-1}$) for selenides (sulphides), the change is more drastic.

The effective masses of electrons and holes at the $K$ point are very stable with respect to the external electric field (see Tab.~\ref{tab:1l}).
Also the band splitting caused by to the spin-orbit effect is not affected by the electric field.
The presence of Stark effect, however, is pronounced and can be observed in the shape of the band structures (see Fig.~\ref{fig:BS_S}).
\begin{table}[h!]
\footnotesize{
\caption{\label{tab:1l} Spin-orbit splitting $\Delta_{\textrm{SO}}$ of the highest occupied valence band and effective masses of electrons and holes at $K$ point versus the electric field of MoWSeS monolayers.}
\centering
\begin{tabular}{c|c|*{3}{c}}
  \hline\hline
 System   & E$_{\textrm{field}}$ & \multicolumn{3}{|c}{at $K$ point} \\ \cline{3-5}
             & (V \AA$^{-1}$)    & $\Delta_{\textrm{SO}}$(meV)  & m$^*_e$/m$_0$ &m$^*_h$/m$_0$ \\ \hline
                       &   0.00     &       150      &    0.450    &  -0.537    \\
  MoS$_2$              &   3.00     &       150      &    0.451    &  -0.546    \\
                       &   4.00     &       149      &    0.451    &  -0.551    \\ \hline
  
                       &   0.00     &       183      &    0.561    &  -0.614    \\
  MoSe$_2$             &   2.00     &       181      &    0.564    &  -0.620    \\
                       &   4.00     &       180      &    0.556    &  -0.630    \\ \hline

                       &   0.00     &       430      &    0.367    &  -0.334    \\
   WS$_2$              &   3.00     &       424      &    0.381    &  -0.342    \\
                       &   4.00     &       415      &    0.393    &  -0.347    \\\hline
                       
                       &   0.00     &       453      &    0.426    &  -0.355    \\
   WSe$_2$             &   2.00     &       443      &    0.435    &  -0.360    \\
                       &   4.00     &       441      &    0.455    &  -0.372    \\ 
                 \hline
\end{tabular}
}
\end{table}

Applying external electric field causes changes especially in the conduction band and other minima, namely between the $\Gamma-M$ and $K-\Gamma$, become close in energy to the CBM.
Therefore, it might be of interest to investigate the electron effective masses for those minima (see Tab.~\ref{tab:em}).
The effective masses of electron at those $k$ points are more sensitive to the external electric field than at the $K$ point and reduce (increase) significantly for the $\Gamma-M$ ($K-\Gamma$).
At zero fields, they are larger than the corresponding values at the $K$ point, though.
Note, we do not report the hole effective masses for other maxima in the valence band (at the $\Gamma$ point), as the bands are very flat and the values become very large and meaningless.
\begin{table}[h!]
\footnotesize{
\caption{\label{tab:em} Effective masses of electrons for minima along the $\Gamma-M$ and $K-\Gamma$ paths of MoWSeS monolayers for selected electric field strengths.}
\centering
\begin{tabular}{c|c|cc*{4}{c}}
  \hline\hline
 System   & E$_{\textrm{field}}$ &  \multicolumn{1}{c}{m$^*_e$/m$_0$}   \\ \cline{3-4}
              & (V \AA$^{-1}$) &  \multicolumn{1}{c|}{CBM ($\Gamma-M$)} & CBM ($K-\Gamma$)  \\ \hline
    MoS$_2$   &    0.00        &     0.956      &    0.599       \\
              &    3.00        &     0.676      &    1.259        \\ \hline
   MoSe$_2$   &    0.00        &     0.776      &    0.5417       \\
              &    2.50        &     0.653      &    1.305        \\ \hline
   WS$_2$     &    0.00        &     0.886      &    0.541        \\
              &    3.00        &     0.619      &    0.869         \\ \hline
    WS$_2$    &    0.00        &     0.694      &    0.436        \\
              &    2.50        &     0.565      &    1.175        \\ \hline
\end{tabular}
}
\end{table}

\begin{figure}[h!]
\begin{center}
\includegraphics[scale=0.8,clip]{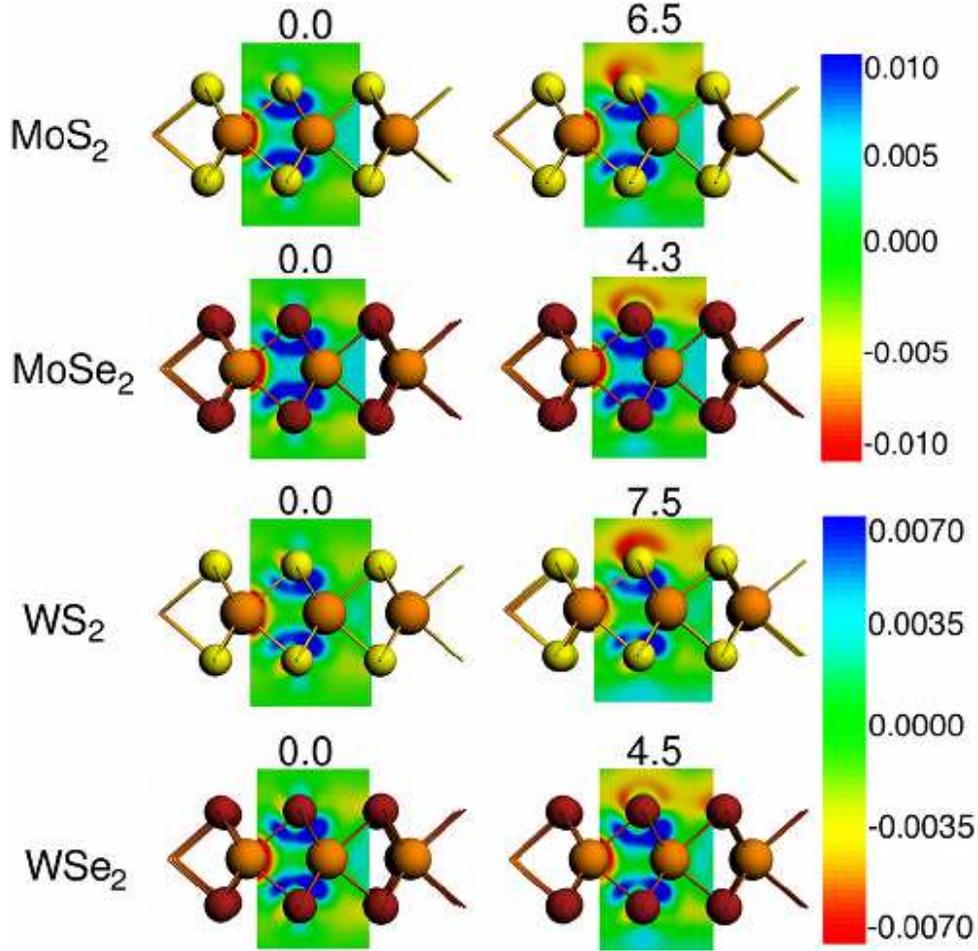}
\caption{\label{fig:Den} Deformation density maps (a.u.) of TX$_2$ systems at zero (left) and finite (right) external electric field (V \AA$^{-1}$).}
\end{center}
\end{figure}

Fig.~\ref{fig:Den} shows the deformation density maps at zero and the critical electric field strengths.
In these plots, the red zone refers to electron depletion, while the blue colour corresponds to an excess of electrons.
In the equilibrium, there is a uniform charge distribution, symmetric with respect to the transition-metal layer for all MoWSeS monolayers.
Once the electric field reaches a critical strength, that is at the point of semiconductor-metal transition, we observe a strong polarization of the system, indicated by the red colour that starts to dominate in the upper chalcogen layer, while the blue colour appears in the lower chalcogen layer.
This means that a strong polarization occurs normal to the basal plane, inducing a dipole moment, which leads to the reduced band gaps. 

In conclusion, we show that MoWSeS monolayers TX$_2$, T = Mo, W, X = S, Se are very stable with respect to external electric fields that are common when applying a gate voltage.
Only at very strong fields, exceeding 2 V \AA$^{-1}$, effects on the electronic structure become notable.
Those include a Stark effect and the change of electronic structure, leading to a transition from direct to indirect band gap in the monolayers.
Even though the band structures, in particular the conduction bands, change due to the Stark effect, quantities that dominate the electronic properties of the materials such as band gap, effective masses of electrons and holes, and the value of the spin-orbit splitting are unaffected by the external fields.

\section{Acknowledgements}
Financial support by Deutsche Forschungsgemeinschaft (DFG, HE 3543/17-1) and the European Commission through the Initial Training Network (ITN) MoWSeS (GA FP7-PEOPLE-2012-ITN) and the Industrial Academic Partnership Pathways (IAPP) QUASINANO (GA FP7-PEOPLE-2009-IAPP) is acknowledged.

\end{document}